\documentclass[sigconf]{acmart}

\usepackage{graphicx}
\usepackage{epstopdf}
\usepackage{booktabs} 
 \usepackage{amsmath}
\usepackage{amsmath}
\usepackage{amsthm}

\usepackage{amsmath}
\usepackage{bm}
\usepackage{algorithm}
\usepackage{algorithmic}
\usepackage{amsmath}
\usepackage{multirow}
\usepackage{subfigure}
\usepackage{epsfig}
\usepackage{epstopdf}
\usepackage{xspace}
\usepackage[normalem]{ulem}
\usepackage{amsmath}
\usepackage{dsfont}
\usepackage{microtype}
\usepackage{graphics}
\usepackage{bbm}

\AtBeginDocument{%
  \providecommand\BibTeX{{%
    \normalfont B\kern-0.5em{\scshape i\kern-0.25em b}\kern-0.8em\TeX}}}

\newcommand{\paratitle}[1]{\vspace{1.5ex}\noindent\textbf{#1}}

\newcommand{\ignore}[1]{}

\copyrightyear{2023} 
\acmYear{2023} 
\setcopyright{acmlicensed}\acmConference[CIKM '23]{Proceedings of the 32nd ACM International Conference on Information and Knowledge Management}{October 21--25, 2023}{Birmingham, United Kingdom}
\acmBooktitle{Proceedings of the 32nd ACM International Conference on Information and Knowledge Management (CIKM '23), October 21--25, 2023, Birmingham, United Kingdom}
\acmPrice{15.00}
\acmDOI{10.1145/3583780.3615498}
\acmISBN{979-8-4007-0124-5/23/10}

\begin{document}

\title{RUEL: Retrieval-Augmented User Representation with
Edge Browser Logs for Sequential Recommendation}

\author{Ning Wu}
\affiliation{\institution{Microsoft STCA}
            \country{China}}
\email{wuning@microsoft.com}


\author{Ming Gong}
\authornote{Daxin Jiang is the corresponding author.}
\affiliation{\institution{Microsoft STCA}
            \country{China}}
\email{migon@microsoft.com}

\author{Linjun Shou}
\affiliation{\institution{Microsoft STCA}
            \country{China}}
\email{lisho@microsoft.com}

\author{Jian Pei}
\affiliation{\institution{Duke University}
            \country{USA}}
\email{j.pei@duke.edu}

\author{Daxin Jiang}
\affiliation{\institution{Microsoft STCA}
            \country{China}}
\email{djiang@microsoft.com}



\begin{abstract}
 
Online recommender systems (RS) aim to match user needs with the vast amount of resources available on various platforms. A key challenge is to model user preferences accurately under the condition of data sparsity. To address this challenge, some methods have leveraged external user behavior data from multiple platforms to enrich user representation. However, all of these methods require a consistent user ID across platforms and ignore the information from similar users. In this study, we propose RUEL, a novel retrieval-based sequential recommender that can effectively incorporate external anonymous user behavior data from Edge browser logs to enhance recommendation. We first collect and preprocess a large volume of Edge browser logs over a one-year period and link them to target entities that correspond to candidate items in recommendation datasets. We then design a contrastive learning framework with a momentum encoder and a memory bank to retrieve the most relevant and diverse browsing sequences from the full browsing log based on the semantic similarity between user representations. After retrieval, we apply an item-level attentive selector to filter out noisy items and generate refined sequence embeddings for the final predictor. RUEL is the first method that connects user browsing data with typical recommendation datasets and can be generalized to various recommendation scenarios and datasets. We conduct extensive experiments on four real datasets for sequential recommendation tasks and demonstrate that RUEL significantly outperforms state-of-the-art baselines. We also conduct ablation studies and qualitative analysis to validate the effectiveness of each component of RUEL and provide additional insights into our method.

\end{abstract}

\ccsdesc[500]{Information systems~Personalization}


\keywords{User Browsing Log, Sequential Recommendation, Dense Retrieval, Contrastive Learning}


\maketitle

\section{Introduction}
 

Recommender systems (RS) play a vital role in various online applications by matching user needs with a large number of candidates (called items). A key challenge for RS is to accurately characterize and understand users’ interests and preferences, which are often sparse and noisy in user-item interactions. To enhance the recommendation performance, recent studies have leveraged various types of side information to enrich the representation of users and items, such as spatial-temporal information~\cite{lian2020geography}, user/item attributes~\cite{zhou2020s3}, knowledge graph ~\cite{wang2019kgat, wang2018ripplenet} and cross-domain behavior~\cite{wu2019neural, yao2021user, si2022model}.


\begin{figure}[t]
 \includegraphics[width=1.0\linewidth]{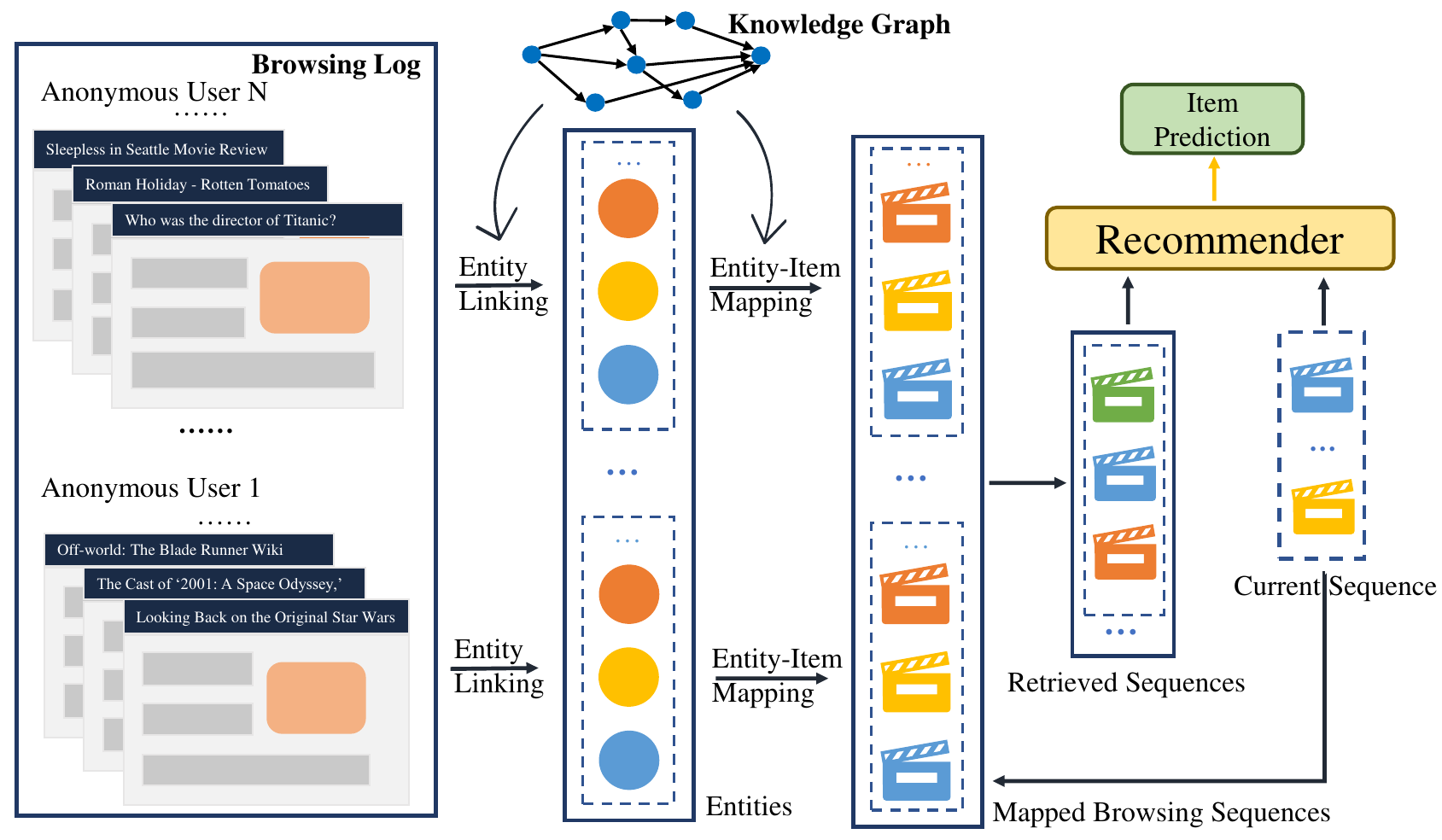}\vspace{-.3cm}
 \caption{Illustration of the overall procedure on movie recommendation. Webpages browsed by users are first linked to entities by entity linking technology. Then these entities are mapped to items based on item name, publish years and etc. Finally, a well-trained retriever will search for useful browsing sequence and feed them into recommender. }
 \label{intro-framework}\vspace{-.3cm}
\end{figure}


Web browsing behavior data can reveal user interests and preferences that are useful for recommender systems. For example, many online users search for information they need using web browsers such as Chrome and Edge. The browsing data and search queries collected by these browsers can cover a large and diverse user population. However, most existing methods for leveraging browsing data in recommender systems only focus on the users who are also in the recommendation datasets ~\cite{wu2019neural, yao2021user, si2022model}. This limits the applicability and effectiveness of these methods for two reasons. First, they require a unified user identifier across multiple domains, which is not always available or feasible. Second, they ignore the potential benefits of using the browsing data of other anonymous users who may have similar interests to the target user. Motivated by the recent advances in open-domain question answering using dense retrieval~\cite{qu2021rocketqa, ren2021rocketqav2} and some knowledge-based recommendation methods~\cite{wang2019kgat, wang2018ripplenet}, we aim to bridge the gap between the Edge browsing log and entity-related recommendation tasks such as movie, book and music recommendations. The overall procedure is illustrated in Figure~\ref{intro-framework}. However, the browsing log is very noisy and sparse, and many webpages may not be relevant to the candidate items. The key challenge is how to select and represent useful browsing data in the presence of noise.
To address this challenge, we propose a novel retrieval-augmented sequential recommender based on Edge browsing log, named RUEL, which stands for \uline{R}etrieval-\uline{A}ugmented \uline{U}ser Representation with \uline{E}dge \uline{B}rowser Logs for Sequential \uline{R}ecommendation. First, we apply an augmented contrastive learning framework to the encoder that takes the current user behavior sequence as input. We generate two augmented views from each user behavior sequence and feed them into a transformer encoder. The objective of contrastive learning is to maximize the agreement between the two views from the same sequence. This has two advantages: (1) it improves the embedding space and reduces anisotropy~\cite{qiu2022contrastive}, and (2) it enhances the model robustness to noise by using data augmentation techniques~\cite{xie2020contrastive, liu2021contrastive}. Moreover, we construct a momentum encoder and a memory bank for browsing sequences. The momentum encoder encodes browsing sequences and stores their embeddings in the memory bank. All embeddings in the memory bank are used as negative samples in contrastive learning. The encoder is trained to distinguish the augmented original user behavior sequence from a large number of browsing sequences. Finally, we convert all browsing sequences into a retrieval index using the retriever for fast retrieval. At inference time, we retrieve top k browsing sequences from the index for each user, and assign them different weights by an item-level attentive selector. The predictor aggregates multiple weighted sequence embeddings by attention mechanism, and predicts the target item.
In summary, our main contributions are as follows.
\begin{itemize}
    \item We propose a novel approach to mine useful patterns from anonymous browsing data to improve recommender systems. We bridge the gap between anonymous webpage browsing data and various recommendation tasks.
    \item We use a momentum contrastive learning framework on user behavior sequences and anonymous browsing sequences to train a powerful retriever, and design an attentive selector to generate fine-grained weights for each retrieved sequence.
    \item We conduct extensive experiments on four real-world datasets to demonstrate the effectiveness and robustness of our proposed approach.
\end{itemize}

\section{RELATED WORK}

Our work is related to the following research directions.

\paratitle{Sequential Recommendation.}  
By modeling high-order dependency between between each items, sequential recommenders aims to recommend appropriate items to users. Previous \cite{rendle2010factorizing,he2016fusing,cai2017spmc} efforts utilize MCs to identify first-order transition relationships. Subsequently, with the rapid growth of deep learning techniques, numerous works \cite{hidasi2015session,quadrana2017personalizing,li2017neural,hidasi2018recurrent,hidasi2016parallel,jannach2017recurrent} has been developed to apply deep models to SR. GRU4Rec \cite{hidasi2015session} first introduces RNN to the SR task, taking into account practical features of the task and a number of modifications to standard RNNs, such as a ranking loss function. Caser \cite{tang2018personalized} uses convolutional filters to learn sequential patterns as local features of the image. In addition, influenced by the success of the attention mechanism and Transformers in other domains \cite{vaswani2017attention,devlin2018bert,wang2018non,yang2019xlnet,brown2020language,raffel2020exploring,liu2019roberta}, SASRec \cite{sasrec} has achieved significant performance improvements by first utilizing self-attention to represent the interplay of past interactions. Then, BERT4Rec \cite{bert4rec} models user behavior sequences using deep bidirectional self-attention by adopting the Cloze objective to SR.  ReDA\cite{bian2022relevant} generates relevant and diverse augmentation by the related information from similar users.

\paratitle{Contrastive Learning.}
Self-supervised learning (SSL) has attracted widespread attention in past several years, \cite{he2020momentum,grill2020bootstrap,chen2020simple,lan2019albert,oord2018representation,qiu2020gcc}. MoCo \cite{he2020momentum} design a momentum mechanism to enhance memory bank mechanism and SimCLR \cite{chen2020simple} proposes a simple contrastive learning framework without memory bank and any specialized architectures. SimSiam \cite{chen2021exploring} is a conclusive work on doing contrastive learning with convolutional neural networks.  S3-Rec \cite{zhou2020s3} is built on self-attention architecture, and proposes to use attribute information to produce self-supervision signals and augment data representations. SGL \cite{wu2021self} generates multiple views of a subgraph, and maximizes the agreement between different views of the same node in two subgraphs CLS4Rec \cite{xie2020contrastive}, DuoRec \cite{qiu2022contrastive}, MMInfoRec \cite{qiu2021memory}, CoSeRec \cite{liu2021contrastive} and ContraRec \cite{wang2022sequential} proposes to utilize contrastive learning to empower sequential recommendation.  

\begin{table}[!t]
  \centering
  \caption{Interaction information statistics for browsing datasets. We present the total browsing webpage numbers of each dataset after preprocessing of three stages. }
    \begin{tabular}{l|cccc}
    \toprule
    \textbf{Dataset} & \textbf{Raw} & \textbf{Stage 1} & \textbf{Stage 2} & \textbf{Stage 3} \\
    \midrule
    \textbf{ML-1m} & 32b  & 871m & 537m   & 22m \\
    \textbf{ML-20m} & 32b  & 871m & 537m  & 95m \\
    \textbf{Amazon-Book} & 32b  & 108m  & 75m  & 27m \\
    \textbf{Last FM} & 32b  & 978m  &  538m  & 14m \\
    \bottomrule
    \end{tabular}%
  \label{tab:browsing_stat}%
\end{table}%

\section{Preprocessing}



We preprocess the browsing data in three stages: entity linking, session segmentation, and item alignment. In the first stage, we use Microsoft Satori to link webpages to entities based on their title and main text. We build a billion-level webpage-entity dictionary by retrieving and ranking candidate entities for each webpage using BM25 and a roberta-based ranking model. The best entity with a ranking score model above 0.9 is selected. With the webpage-entity dictionary, we filter raw browsing data by only keeping webpages that are linked to Movie/Book/Artist entities. In the second stage, we split the browsing log of each user into sequences with lengths greater than 4 using a 4-hour time interval. Then we use side information in the dataset to match these entities to candidate items in three datasets~\cite{kb4rec}. For Movie-lens dataset, we compare the release year and movie name. For Amazon-book dataset, we use the author name, and book name. For the Last FM dataset, we use the artist's name. We choose the top-1 candidate after verification for each item. Table \ref{tab:browsing_stat} shows the statistics of the browsing data after each stage.

\section{TASK Formulation}

We consider a sequential recommendation task with a set of users $\mathcal{U}=\{u_1,\cdots, u_{|\mathcal{U}|}\}$ and a set of items $\mathcal{V}=\{{v_1,\cdots, v_{|\mathcal{V}|}}\}$. 
The user-item interaction matrix $Y=\{y_{uv}|u\in \mathcal{U}, v\in \mathcal{V} \}$ captures the implicit feedback of users, where $y_{uv}=1$ indicates that user $u$ has interacted with item $v$, and $y_{uv}=0$ otherwise. The interaction can be any type of behavior such as clicking, watching, browsing, etc.  For each user $u$, we can also obtain the interaction sequence $s_u = (v_{1},...,v_{j},...,v_{l_u})$, where $s_u \in \mathcal{C}$, $v_{j}$ is the item that $u$ has interacted with at time step $j$, and $l_u$  is the length of the interaction history for user $u$.

Moreover, we assume that we have access to a large amount of webpage browsing data $\mathcal{D}$ from Edge browser, which consists of numerous webpage sequences $s_{i}^r$, where $i$ denotes the $i$-th browsing session of anonymous users from $\mathcal{D}$. 
Based on these definitions, we formulate the retrieval-augmented recommendation problem as follows. Given the interaction sequence $s_u = (v_{1},...,v_{t},...,v_{l_u})$ of user $u$, and the webpage browsing data $\mathcal{D}$ from Edge browser, our goal is to predict the next item $v_t$ that user $u$ will interact with.

\section{The RUEL Model}

\begin{figure}[t]
 \includegraphics[width=1.03\linewidth]{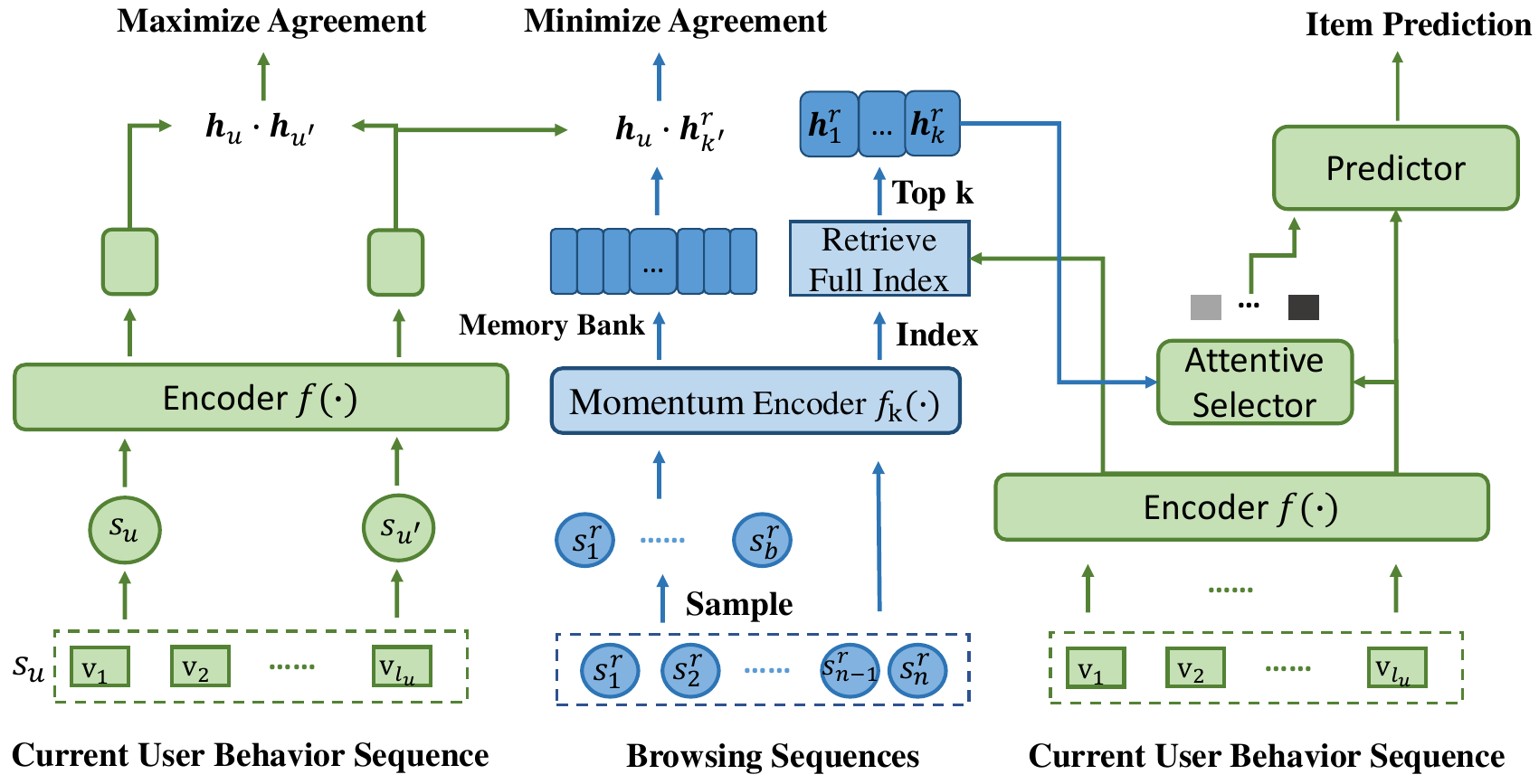}\vspace{-.3cm}
 \caption{The overall architecture of the RUEL model. $f(\cdot)$ is trained to maximize agreement between positive pairs, and discriminate positive pairs from enormous negative embeddings generated by momentum encoder $f_k(\cdot)$. Blue part depicts browsing sequences are sent into $f_k(\cdot)$, and converted into negative embeddings in memory bank. After first stage training, all browsing sequences are encoded into embeddings to construct full retrieval index.
 }
 \label{fig-framework}\vspace{-.3cm}
\end{figure}

In the section, we present the proposed \emph{RUEL: Retrieval-Augmented User Representation with Edge Browser Logs for Sequential Recommendation}. Figure \ref{fig-framework} illustrates our model framework, which consists of three main components: 1) Contrastive learning with multiple data augmentation strategies; 2) Momentum encoder and memory bank. They help the encoder learn to discriminate positive sample from enormous browsing negative samples; and 3) prediction module, which consists of an attentive selector on item-level and a predictor combines retrieved information and current item sequence to predict the next item.



\subsection{Transformer-based Recommender}
Following previous work~\cite{qiu2022contrastive, bert4rec, sasrec}, we also choose transformer encoder~\cite{vaswani2017attention} as modeling tool $f(\cdot)$ of input sequence $s_u$, which has been widely applied to numerous CV and NLP tasks.  For a browsing sequence $s_i$, we use $\bm{h}_u$ to denote its representation generated by transformer encoder. Furthermore, we use $\bm{h}_{u,j}$ to denote embedding of $j$-th item in $u$-th user in $\mathcal{C}$, and use $\bm{h}_{i,j}^r$  to denote embedding of $j$-th item in $i$-th browsing session in $\mathcal{D}$.

\subsection{Contrastive Learning for Sequential Modeling} 
We first introduce the data augmentation methods used in this paper. Then, we explain the technical aspects of momentum contrastive learning on browsing sequences.

\subsubsection{Sequential Augmentation}
Given current items sequence $s_u$, we directly introduce some typical sequence-based augmentations~\cite{liu2021contrastive}. We randomly select an operator from \textit{Mask}, \textit{Crop}, \textit{Reorder} in each augmentation.


\begin{itemize}
    \item \textbf{Mask}(M).  The item masking approach \cite{zhou2020s3,cheng2021learning,xie2020contrastive} randomly replaces each token with special token \textit{[Mask]} by a probability $\gamma \in (0,1)$.  This augmentation strategy can be formulated as: $s_{u^{'}}^M=[v_1, v_\text{[Mask]},\cdots,v_{l_u}]$.

    \item \textbf{Reorder}(R). Different item orders may reflect the same user interests\cite{cheng2021learning,xie2020contrastive}. We can shuffle a continuous sub-sequence $[v_{r+1},\cdots,v_{r+l_r}]$ to $[v_r^{'},v_{r+1}^{'},\cdots,v_{r+l_r}^{'}]$, where $l_r=\mu*k$ is the sub-sequence length and $\mu \in (0,1)$.

    \item \textbf{Crop}(C). Given a current behavior sequence $s_u$, we randomly select a sub-sequence $s_{u^{'}}^{C}=[v_{c+1},\cdots,v_{c+l_c}]$ with length $l_c= \eta* l_u $, where $\eta \in (0,1)$ is a length hyper-parameter.
\end{itemize}




\subsubsection{Momentum Contrastive Learning}
Moreover, we adopt momentum contrastive learning, which uses a large memory bank with many negative examples, and maintains a consistent encoder for generating negative embeddings with a momentum-based update mechanism. The memory bank is updated in a \emph{FIFO}~(First-In-First-Out) fashion. At each iteration, the representation $\bm{h}_i^{r}$ is generated by a key encoder $f_k(\cdot)$ that has the same structure as $f(\cdot)$. Then $\bm{h}_i^{r}$ is added to $\bm{M}$, and the oldest representation in the memory bank is removed. Notably, our memory bank $\bm{M}$ stores the embeddings of browsing sequences as negative examples to enhance the discriminative power of the query encoder.  Using a queue-based memory bank allows us to increase the size of negative samples in each loss computation, but it prevents us from updating the key encoder $f_k(\cdot)$ via back-propagation (the gradient is disconnected for all samples in the queue). Following~\cite{he2020momentum}, the parameters of the key encoder $\theta_k$ are updated by following those of the query encoder $\theta_q$ with a momentum coefficient $m$, resulting in stable representations for similar sequences. Formally, we have:

\begin{equation}
\label{eq:7}
\theta_k \gets m\theta_k+(1-m)\theta_q,
\end{equation}
where $m \in [0,1)$ is a hyper-parameter. Back-propagation only modifies $\theta_q$. The momentum update in Equation \ref{eq:7} makes $\theta_k$ evolve more smoothly than $\theta_q$. We simply set $m$ to 0.999 following~\cite{he2020momentum}.
Based on the memory bank, we optimize the sequence representation with a momentum contrastive loss function as:

\begin{equation}
\label{eq:8}
    \mathcal{L}_{s_u,s_{u^{'}}}=-\log{\frac{\exp(\bm{h}_u\cdot \bm{h}_{u^{'}}/{\tau})}{\sum_{i=1}^{2N} \mathds{1}_{[u \neq i]} \exp(\bm{h}_u\cdot \bm{h}_{i}/{\tau})  + \sum_{k^{'}=1}^K  \exp(\bm{h}_u\cdot \bm{h}_{k^{'}}^r/{\tau}) }},
\end{equation}
where $\tau$ is a temperature hyper-parameter, $K$ is the size of memory bank. This loss is equivalent to minimizing the negative log likelihood of a $(2N+K)$-way softmax classifier that tries to distinguish j-th sequence from other sequences in the batch and many browsing sequences in the memory bank.

\begin{equation}
    \mathcal{L}_{CTS}=\sum_{u \in \mathcal{D}} \mathcal{L}_{s_{u}, s_{u^{'}}}
\end{equation}

In above equation, $s_u$ denotes each behavior sequence  in dataset $\mathcal{C}$, and we use $s_{u^{'}}$ to denote augmented view of it. 

\subsection{Retrieval-augmented Sequential Modeling} In this section, we present our retrieval-augmented sequential recommendation model, which uses the observed user behavior session data $s_u$ to retrieve top-k browsing sequences $\{s_1^r, s_2^r, ..., s_k^r\}$ and use them as additional context for sequential recommendation. As shown on the right of Figure 2, after top-k sessions are retrieved from full index, we perform fine-grained attentive selection on item level. For each retrieved behavior sequence, its final representation is computed by a weighted sum of its item representations. Then, a predictor takes both weighted retrieved sequence embedding and current user sequence embedding as input, and predicts target item.

\subsubsection{Session-level Retrieval}
We use transformer encoders $f_k(\cdot)$ to build index of retriever. For each user in browsing data, their records are split into different sessions by time interval as stated above. The encoder $f_k(\cdot)$ takes these sessions as input, and produces dense embedding $\bm{h}_i^r$ for each browsing sequence $s_i^r$ in $\mathcal{D}$. Then a relevance score $f(s_u,s_i^r )=\bm{h}_u^T \bm{h}_i^r$ is calculated between dense session vector $\bm{h}_u$ of user u and browsing sequence vector $\bm{h}_i^r$  by inner product. Thus, we can employ Maximum Inner Product Search (MIPS) algorithms to find the approximate top k sessions$\{s_1^r, s_2^r, ..., s_k^r\}$, using running time and storage space that scale sub-linearly with the number of browsing sequences. We use faiss\footnote{https://github.com/facebookresearch/faiss} to implement above-mentioned retrival procedures.

\subsubsection{Attentive Selection}
After top $k$ sessions are retrieved from numerous browsing data, we aim to reduce the impact of noisy items on prediction. Therefore we apply an item-level attention mechanism to compute attention weight for each item in retrieved sessions.

\begin{align}
\alpha_{u}^j &=\frac{\exp((\bm{W}_1 \bm{h}_u  + \bm{W}_2 \bm{h}_{i,j}^r))}{\sum_{j}^{l_i}\exp(\bm{W}_1 \bm{h}_u + \bm{W}_2 \bm{h}_{i,j}^r)} \\
\bm{o}_i &=\sum_{j \in \{1 \cdots l_i\}} \alpha_{u}^j \bm{h}_{i,j}^r 
\end{align}
where $l_i$ denotes the length of $s_i^r$, $\bm{h}_{i,j}^r$ denotes the output hidden state of $j$-th item in $i$-th  retrieved browsing sequence $s_i^r$, and $\bm{o}_i$ denotes the weighted embedding of sequence $s_i^r$. $\bm{W}_1$ and $\bm{W}_2$ are learnable parameters. $\alpha_u^j$ represents the attention weight between $s_u$ and the $j$-th items in $i$-th retrieved browsing behavior sequence $s_i^r$. After obtaining $\bm{o}_i$ for each retrieved sequence, we directly use the retrieve score $f(s_u, s_i^r)$ as weight of $\bm{o}_i$ to compute the final context vector $\bm{o}$.

\begin{equation}
\bm{o} =\sum_{i \in \{1...k\}} f(s_u, s_i^r) \bm{o}_i
\end{equation} 
where $f(s_u, s_i)$ is the dot product relevance score generated by retriever. In summary, we select top-k retrieved browsing sequences from both sequence level and item level, and obtain a comprehensive context vector $\bm{o}$ to enhance the predictor.

\subsubsection{Prediction and Loss Function}
Given the aggregated context vector $\bm{o}$, we use a two-layer multilayer perceptron~(MLP) to capture the high-level interaction between $\bm{o}$ and $\bm{h}_u$.
\begin{equation}
p(v_{t}|s_u, \mathcal{D})=\frac{\exp(\bm{w}_{t}^T \text{MLP}(\bm{h}_{u} \mathbin\Vert  \bm{o}))}{\sum_{j=1}^N\exp(\bm{w}_{j}^T\text{MLP}(\bm{h}_{u} \mathbin\Vert \bm{o}))}
\end{equation}
where $\bm{w}_j^T$ is a learnable embedding for item $j$, and $\mathcal{D}$ denotes the full set of browsing data. The final  loss function is cross entropy, which can be written as:

\begin{equation}
    \mathcal{L}_{CF}=\sum_{s_u \in \mathcal{C}} -\log p(v_{t}|s_u, \mathcal{D}),
\end{equation}
where $\mathcal{C}$ denotes the set of all user behavior sequences. $v_t$ denotes the target item of behavior sequence $s_u$.

\subsubsection{Optimization and Training Strategy.}
We adopt a two-stage training strategy. In the first stage, we train the transformer encoder $f(\cdot)$ by minimizing the contrastive loss in Equation~\ref{eq:8}, and update the momentum encoder $f_k(\cdot)$ by Equation~\ref{eq:7}. Then we construct a fixed browsing sequence index using $f_k(\cdot)$. In the second stage, we jointly optimize the contrastive loss and the cross entropy loss.

\begin{equation}
    \mathcal{L}=\mathcal{L}_{CTS} + \mathcal{L}_{CF},
\end{equation}

Unlike some previous work~\cite{guu2020retrieval}, which rely on a reward to guide the gradient of the retriever and constantly refresh the index with the updated retriever, we use a fixed browsing data index, which allows us to directly optimize the transformer encoder $f(\cdot)$ in the item prediction task by gradient backpropagation. By sharing the encoder in both momentum contrastive learning and item prediction task, the transformer encoder $f(\cdot)$ will also benefit from the contrastive learning, because a good embedding space is also conducive to item prediction~\cite{qiu2022contrastive, liu2021contrastive}.

\section{EXPERIMENTS}

In this section, we first set up the experiments, and then present the performance comparison and analysis. 


\subsection{Experimental Setup}

\paratitle{Edge Browser Log Mining.}
We present the details of entity linking and entity-item mapping in Section 4. Then, we can transform the user browsing log into a sequence of entities or items.  Due to the limitation of resources, we only use user browsing log of Edge browser in 2021. We collect more than 32 billion records and convert them to browsing sequences for four datasets. Table~\ref{tab:inter_stat} shows detailed information.

\begin{table}[!t]
  \centering
  \caption{Interaction information statistics for datasets. We use \emph{Inter} to denote interaction. }
    \begin{tabular}{l|cccc}
    \toprule
    \textbf{Dataset} & \textbf{\#User} & \textbf{\#Item} & \textbf{\#Inter} & \textbf{\#Inter$_{\text{Browsing}}$} \\
    \midrule
    \textbf{ML-1m} & 6,040  & 3,416 & 1,000,000   & 22,704,450 \\
    \textbf{ML-20m} & 138,493  & 26,744 & 20,000,000  & 95,148,210 \\
    \textbf{Amazon-Book} & 281,428  & 13,044  & 3,500,000  & 27,439,633 \\
    \textbf{Last FM} & 1,090  & 3,646  &  52,538  & 14,247,218 \\
    \bottomrule
    \end{tabular}%
  \label{tab:inter_stat}%
\end{table}%

\paratitle{Evaluation Metrics.}
In this paper, we use Normalized Discounted Cumulative Gain (NDCG) and Hit Ratio (HR) as metrics~\cite{lim2015personalized, kurashima2010travel, cui2018personalized}. We set k=5 and k=10 in experiments. 




\paratitle{Task Setting.} 
For data preprocessing, we follow the common practice in previous work. To ensure the quality of the dataset, we only keep users with at least five interactions following previous practice. To evaluate the sequential recommendation models, we adopt the leave-one-out evaluation (i.e., next item recommendation) task. For each user, we hold out the last item of the behavior sequence as the test data and use the item just before the last as the validation set. The remaining items are used for training. We implemented our baselines based on the RecBole \cite{zhao2021recbole} framework. All baseline settings and training strategies refer to the original authors' implementation and further tune the parameters based on it. For fairness, the embedding size, layers number and heads number of all transformer-based models are fixed to the widely used 128, 2 and 2. The size of memory bank K is set as 8096 directly. We use the Adam optimizer \cite{Adam} to optimize all models. Moreover, all models are trained for 500 epochs, and the early stopping strategy is applied, $\textit{i.e}$., premature stopping if ndcg@10 on the validation set does not increase for 20 successive epochs.


\begin{table}[t] \small
\caption{Performance comparison using four metrics on four datasets. All the results are better with larger values. With paired $t$-test, the improvement of the RUEL over the best baselines is significant at the level of 0.05.}
\newcommand{\re}{$_{re-impl}$}
\newcommand{\tu}{$_{tuned}$}
  \tabcolsep=0.05cm
  \resizebox{1.02\columnwidth}{!}{%
\begin{tabular}{c|l|ccccccccc}

\toprule
\multicolumn{1}{c|}{Datasets} & Metrics 
                                & GRU4Rec                          & SASRec              & Caser              & BERT4Rec      & CL4Rec & CoSeRec        & RUEL & \%Imp.               \\
\midrule \multirow{4}{*}{\shortstack{LF}}

& H@5         &	0.0301&	0.0416&	0.0385&	0.0401&	0.0447&	\uline{0.0484}&	\textbf{0.0526*} & 8.68\% 
           \\ 
& H@10         &	0.0509&	0.0615&	0.0582&	0.0598&	0.0751&	\uline{0.0778}&	\textbf{0.0831*} & 6.81\%
           \\ 

& N@5        &	0.0218&	0.0256&	0.0249&	0.022	&0.0326&	\uline{0.0332}&	\textbf{0.0366*} & 10.24\%
         \\ 

& N@10       &	0.0258	&0.0319&	0.0285&	0.0294	&0.0442&	\uline{0.0450}&	\textbf{0.0492*}
 & 9.33\%        \\

\midrule \multirow{4}{*}{\shortstack{AB}}
& H@5         &	0.0614&	0.0975&	0.0795&	0.0741&	0.1086&	\uline{0.1305}	&\textbf{0.1422*}
& 8.96\%      \\ 
& H@10        &	0.1452&	0.1806&	0.1504&	0.1331&	0.1951&	\uline{0.2115}&	\textbf{0.2220*} & 4.96\%
            \\ 

& N@5       &	0.0341&	0.0614&	0.0351&	0.0404&	0.0693&	\uline{0.0780}&	\textbf{0.0862*} & 10.51\%
  \\ 

& N@10      &	0.066&	0.0851&	0.0634&	0.0681&	0.0995&	\uline{0.1013}&	\textbf{0.1106*} & 9.18\%
       \\

\midrule \multirow{4}{*}{\shortstack{ML-1M}}

& H@5                   & 0.0763           & 0.1087           & 0.0816           & 0.0733           & 0.1147             & \uline{0.1275}                     & \textbf{0.1403*} & 10.04\%     \\ 
& H@10                  & 0.1658           & 0.1904           & 0.1593           & 0.1323           & 0.1975           & \uline{0.2043}                     & \textbf{0.2231*} & 9.20\%       \\ 

& N@5            & 0.0385           & 0.0638           & 0.0372           & 0.0432           & 0.0662      & \uline{0.0715}                     & \textbf{0.0804*}  &12.44\%          \\ 

& N@10          & 0.0671           & 0.0910           & 0.0624           & 0.0619           & 0.0928      & \uline{0.0978}                    & \textbf{0.1046*} & 6.95\%           \\

\midrule \multirow{4}{*}{\shortstack{ML-20M}} 

& H@5       &	0.0825&	0.1135&	0.0915&	0.0801&	0.1285&	\uline{0.1396}&	\textbf{0.1487*} & 6.51\%
       \\ 
& H@10      &	0.1865&	0.2015&	0.1641&	0.1401&	0.2041&	\uline{0.2132}&	\textbf{0.2301*} & 7.93\%
        \\ 

& N@5     &	0.0425&	0.0648&	0.0415&	0.0501&	0.0715&	\uline{0.0785}&	\textbf{0.0866*} & 10.31\%
          \\ 

& N@10    &	0.0731&	0.0992&	0.0698&	0.0665&	0.1051&	\uline{0.1141}&	\textbf{0.1215*} & 6.48\%
            \\

\bottomrule

\end{tabular}
}
\label{tab:main}

\end{table}

\paratitle{Methods to Compare.} We consider the following methods for comparison: GRU4Rec~\cite{jannach2017recurrent}, Caser~\cite{xie2020contrastive}, BERT4Rec~\cite{bert4rec}, CL4Rec~\cite{xie2020contrastive}, CoSeRec~\cite{liu2021contrastive}. Shallow models like BPR-MF~\cite{rendle2012bpr},  NCF~\cite{Wei2012Constructing} and  FPMC~\cite{Wei2012Constructing} are omitted due to length limitation. Among these baselines,  GRU4Rec, Caser and BERT4Rec belongs to typical deep learning based sequential recommender. CL4Rec and CoSeRec are state-of-the-art contrastive  sequential recommender. The parameters in all the models have been optimized using the validation set.

\subsection{Performance Comparison}
Table \ref{tab:main} shows the results of the performance comparison between our RUEL method and the baseline methods on four datasets. We can observe that RUEL consistently achieves the best performance on all datasets. Among the baselines, CL4Rec and CoSeRec are the most competitive ones, especially CoSeRec, which leverages multiple data augmentation strategies to enhance its contrastive learning.  We can also see that the transformer-based recommenders have an advantage over other deep learning architectures. Moreover, we notice that our method has a larger gain on ML-1M than on ML-20M, indicating that our method is more effective on relatively small datasets.

\begin{table}[!htb]
  \centering
  \caption{Ablation analysis on the Amazon-Book and ML-1m datasets. }
  \setlength{\tabcolsep}{1.0mm}{
  \resizebox{1.02\columnwidth}{!}{
    \begin{tabular}{l|cc|cc|cc|cc}
    \toprule
    \multicolumn{1}{c|}{\multirow{3}[4]{*}{Arch.}} & \multicolumn{4}{c|}{Amazon-Book} & \multicolumn{4}{c}{ML-1M} \\
\cmidrule{2-9}  & \multicolumn{2}{c|}{Reco} & \multicolumn{2}{c|}{Retrieval}   & \multicolumn{2}{c|}{Reco} & \multicolumn{2}{c}{Retrieval} \\         
\cmidrule{2-9}          & H@10  & N@10  & H@10  & H@20  & H@10  & N@10  & H@10  & H@20  \\
\midrule
    RUEL$_{\neg RA}$  & 0.2164  & 0.1026  &0.2348  & 0.3104  & 0.2044  & 0.0985  & 0.1852  & 0.2494  \\
    RUEL$_{\neg MC}$ & 0.2181  & 0.1095 & 0.6591  & 0.8519 & 0.2203  & 0.1015 & 0.6354 & 0.8354 \\
    RUEL$_{\neg DA}$ & 0.2194  & 0.1101 & 0.7214 & 0.8984 & 0.2206  & 0.1032  & 0.6774  & 0.8694 \\
    RUEL$_{\neg AS}$  & 0.2189  & 0.1084 & -  & -  & 0.2204  & 0.1018 & -  & -  \\
    RUEL & 0.2220  & 0.1106 & 0.7249  & 0.9025 & 0.2231  & 0.1046 & 0.6841  & 0.8751 \\
    \bottomrule
    \end{tabular}}
    }
  \label{tab:abla}
\end{table}

\subsection{Ablation Analysis}

We performed ablation experiments on sequential recommendation and retrieval tasks to examine the contribution of each component of RUEL. The results are presented in Table \ref{tab:abla}. For the retrieval task, we utilized our encoder to generate embeddings for a user behavior sequence and its augmented counterpart. We then evaluated the encoder's ability to retrieve the augmented sequence from numerous browsing sequences using HR@10 and HR@20 metrics. In this table, RUEL denotes our full model with all components. RUEL$_{\neg RA}$ eliminates the retriever and the attentive selector components, and directly infers the target item by employing the output embedding of the transformer encoder $f(\cdot)$. RUEL$_{\neg DA}$ discards all data augmentation strategies. RUEL$_{\neg MC}$ removes the memory bank and the momentum mechanism, and only retains batch negatives in contrastive learning. RUEL$_{\neg AS}$ removes the attentive selector, and simply adopts average pooling to obtain the sequence embedding. The results in Table \ref{tab:abla} indicate that RUEL$_{\neg RA}$ slightly outperforms CoSeRec, which implies that our momentum contrast and memory bank are also advantageous for direct recommendation. RUEL$_{\neg DA}$ has a very similar performance to RUEL. It implies that our method does not depend on manually designed sequence augmentation strategies. Discarding the momentum mechanism and the memory bank also deteriorates the model performance. The model needs to be trained to discriminate the original sequence from numerous browsing sequences. The attentive selector plays a vital role in enhancing the model robustness and effectiveness by applying token-level attention. Eliminating this component will substantially impair the model performance.

\begin{table}[!htb]
  \centering
  \caption{A/B Test on Bing Movie. }
  \setlength{\tabcolsep}{3.0mm}{
    \resizebox{0.75\columnwidth}{!}{%
    \begin{tabular}{l|ccc}
    \toprule
Online Metric         & Region \#1  & Region \#2  & Region \#3 \\
\midrule
    WHR  & +2.9\%  & +2.7\%  & +3.4\%   \\
    CTR & +1.2\%  & +1.3\%  & +1.8\%   \\
    Short-Term DAU & +0.07\%  & +0.11\%  & +0.14\%   \\
    \bottomrule
    \end{tabular}}%
    }
  \label{tab:retr}%
\end{table}%

\subsection{Online A/B Test}

We have deployed our model on Bing desktop search by replacing the existed transformer-based sequential ranker with RUEL. To get a stable conclusion, we observe the online experiment for two weeks. Three common metrics in desktop search systems are used to measure the online performance: WHR~(Weighted Hover Rate), CTR~(Click Through Rate), Short-Term DAU. As the result shown in Table~\ref{tab:retr}, the present method RUEL gets overall improvements on multiple regions in our online A/B test experiment.

\subsection{Parameter Analysis}
\begin{figure}[!h]
\centering
\subfigure[Retrieval number k w.r.t ndcg@10.]{
\begin{minipage}[t]{0.45\linewidth}
\centering
\includegraphics[width=1.0\linewidth]{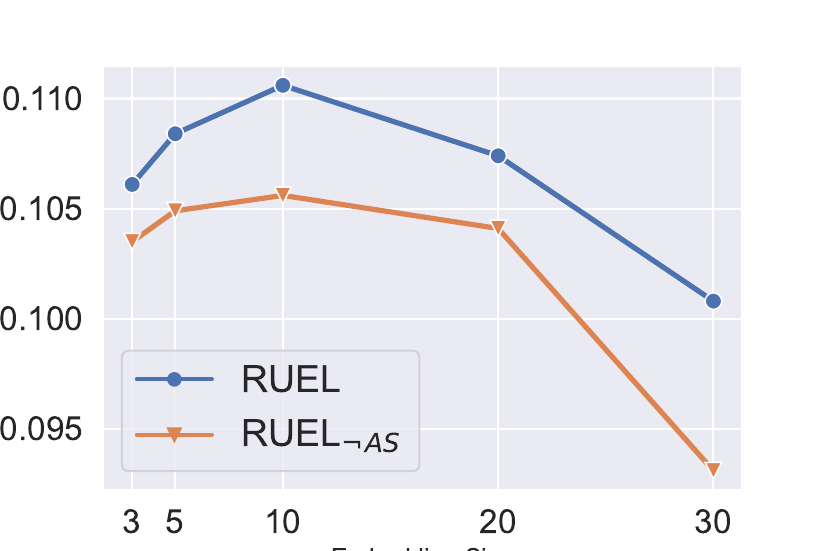}
\end{minipage}\label{fig:case-study-a}
}%
\subfigure[Embedding size w.r.t ndcg@10.]{
\begin{minipage}[t]{0.45\linewidth}
\centering
\includegraphics[width=0.95\linewidth]{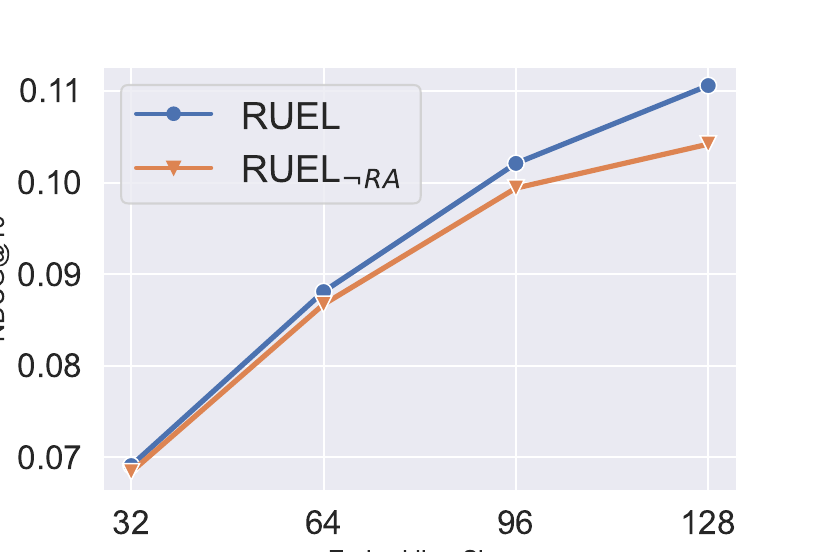}
\end{minipage}\label{fig:case-study-b}
}%
\centering
\caption{Effect of retrieval number and embedding size. } \label{fig:retrieval_number}
\end{figure}

We experiment with RUEL using different $k$ values in [3, 5, 10, 20, 30] and report the results in Figure~\ref{fig:case-study-a}. The metrics peak at $k$=10 on the Amazon-Book dataset and decline afterwards. The attentive selector performs poorly when $k$ is small due to less noise in the top-ranked items. The performance gap between RUEL and RUEL$_\neg AS$ grows as $k$ increases because of more irrelevant items in the retrieved sequences. We also vary the embedding size in [32, 64, 96, 128] and show the results in Figure~\ref{fig:case-study-b}. RUEL consistently outperforms RUEL$_{\neg RA}$ on the Amazon-Book dataset, especially when the embedding size is 128. RUEL benefits more from higher embedding size than RUEL$_{\neg RA}$.

\subsection{Qualitative Analysis}
\begin{figure}[ht!]
 \includegraphics[width=1.0\linewidth]{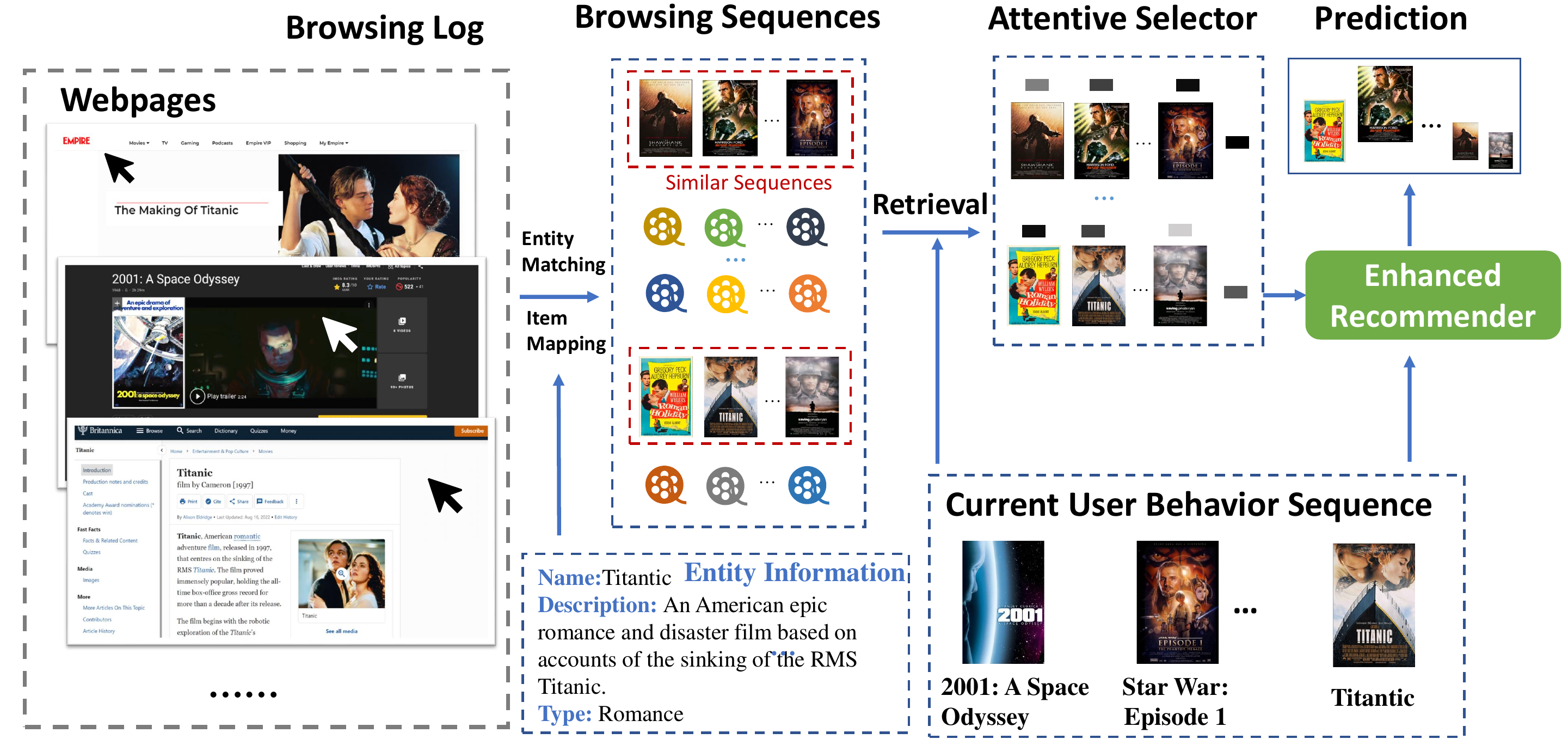}\vspace{-.3cm}
 \caption{Prediction procedure for a sample user in Bing Movie dataset. We use the red frame to denote useful sequences in browsing sequences. For item and sequence-level attention weights, we use the color darkness to indicate attention weights: darker is larger. For predicting results, we use the size to represent predicting probability. 
 }
 \label{fig-case}\vspace{-.3cm}
\end{figure}

We show how RUEL works in Figure~\ref{fig-case}. The red frames are retrieved browsing sequences with popular movies in science fiction and romance genres. The first sequence has \emph{The Shawshank Redemption}, \emph{Star War: Episode 1}, and \emph{The Blade Runner}. The second sequence has \emph{Roman Holiday}, \emph{Titanic} and \emph{Saving Private Ryan}. 
The retriever finds the 10 most similar browsing sequences using the current sequence embedding. The attentive selector assigns a weight to each item in the retrieved sequences. Items that are irrelevant to the user interests, such as \emph{The Shawshank Redemption} in sequence 1 and \emph{Saving Private Ryan} in sequence 2, get lower weights. The item-level and sentence-level weights are combined to produce a context vector. Then the enhanced recommender uses the context embedding and the current sequence embedding to predict the next item. \emph{The Blade Runner} and \emph{Romance Holiday} are the top two candidates because they match the user interests better.

\section{CONCLUSIONS}
This paper proposes a retrieval-augmented sequential recommendation model RUEL to fully utilize Edge browsing information. For an item sequence in the recommendation dataset, it's encoded into a vector, and the vector is used to retrieve similar user behavior sequences from cross-domain behavior. Then retrieved sequences are filtered by an item attentive selector, and refined sequences are used to enhance next item prediction.   In future, we will further explore more heterogeneous cross-domain user behavior data to enhance user modeling.

\bibliographystyle{ACM-Reference-Format}
\bibliography{reference}


\end{document}